\documentclass[conference]{IEEEtran}
\IEEEoverridecommandlockouts

\usepackage{cite}
\usepackage{amsmath,amssymb,amsfonts}
\usepackage{algorithmic}
\usepackage{multirow}
\usepackage{graphicx}
\usepackage{textcomp}
\usepackage{xcolor}
\def\BibTeX{{\rm B\kern-.05em{\sc i\kern-.025em b}\kern-.08em
    T\kern-.1667em\lower.7ex\hbox{E}\kern-.125emX}}
\begin{document}

\title{EmoAttack: Utilizing Emotional Voice Conversion for Speech Backdoor Attacks on Deep Speech Classification Models\\

}



\author{\IEEEauthorblockN{1\textsuperscript{st} Wenhan Yao}
\IEEEauthorblockA{\textit{School of Cyberspace Science} \\
\textit{Xiangtan University}\\
 Xiangtan, China \\
wenhanyao@smail.xtu.edu.cn}
\and
\IEEEauthorblockN{2\textsuperscript{nd} Zedong Xing }
\IEEEauthorblockA{\textit{School of Cyberspace Science} \\
\textit{Xiangtan University}\\
Xiangtan, China \\
202221632998@smail.xtu.edu.cn}
\and
\IEEEauthorblockN{3\textsuperscript{rd} Bicheng Xiong}
\IEEEauthorblockA{\textit{School of Cyberspace Science} \\
\textit{Xiangtan University}\\
Beijing, China \\
202221632999@smail.xtu.edu.cn}
\and
\IEEEauthorblockN{4\textsuperscript{th} Jia Liu}
\IEEEauthorblockA{\textit{School of Software $\&$ Microelectronics} \\
\textit{Peking University}\\
Beijing, China \\
2201120008.pku@vip.163.com}
\and
\IEEEauthorblockN{5\textsuperscript{th} Yongqiang He}
\IEEEauthorblockA{\textit{School of Software $\&$ Microelectronics} \\
\textit{Peking University}\\
Beijing, China \\
heyongqiang@stu.pku.edu.cn}
\and
\IEEEauthorblockN{6\textsuperscript{th} Weiping Wen}
\IEEEauthorblockA{\textit{School of Software $\&$ Microelectronics} \\
\textit{Peking University}\\
Beijing, China \\
weipingwen@pku.edu.cn}
}

\maketitle

\begin{abstract}
Deep speech classification tasks, mainly including keyword spotting and speaker verification, play a crucial role in speech-based human-computer interaction. Recently, the security of these technologies has been demonstrated to be vulnerable to backdoor attacks. Explicitly speaking, speech samples are attacked by noisy disruption and component modification in present triggers. We suggest that speech backdoor attacks can strategically focus on emotion, a higher-level subjective perceptual attribute inherent in speech. Furthermore, we proposed that emotional voice conversion technology can serve as the speech backdoor attack trigger, and the method is called EmoAttack. Based on this, we conducted attack experiments on two speech classification tasks, showcasing that EmoAttack method owns impactful trigger effectiveness and its remarkable attack success rate and accuracy variance. Additionally, the ablation experiments found that speech with intensive emotion is more suitable to be targeted for attacks.

\end{abstract}

\begin{IEEEkeywords}
backdoor attacks, speech classification, emotional voice conversion, speech emotion recognition.
\end{IEEEkeywords}

\section{Introduction}
Speech classification is vital in areas like autonomous driving, smart healthcare, and speaker authentication. It requires extensive data, numerous trainable parameters, and costly resources. Some developers outsource data or model training to third parties to reduce costs and resource demands. Research shows that using third-party platforms for deep neural networks(DNNs) training can introduce security risks\cite{gu2019badnets}, such as data or code poisoning\cite{li2022backdoor,bagdasaryan2021blind}, where attackers embed backdoors into models. These backdoors prompt models to produce incorrect classifications when a specific trigger hides in the inputted samples, making speech classification models vulnerable to such attacks.

Backdoor attacks have been explored earlier in the area of image and text classification. For example, BadNets by Gu et al.\cite{gu2017badnets} and sentences attack by Dai et al.\cite{dai2019backdoor} both involve inserting poisoned samples that cause incorrect model outputs when triggered. These attacks, known as poisoning-label attacks, often modify images using masking, embedding, or color adjustments. However, such techniques may not be effective for speech data. Research shows that speech and image triggers differ due to their distinct physical properties\cite{chen2021mitigating,sun2020natural}. Some speech-trigger methods mimic image-based approaches by inserting noise or specific sounds into speech\cite{koffas2023going,koffas2022can,zhai2021backdoor,shi2022audio,liu2022backdoor,liu2022opportunistic,xin2022natural,luo2022practical}, but these are easily detectable by human ears. To enhance stealthiness, attackers now aim to modify speech components without disrupting them. Ye et al.\cite{ye2023fake} introduced a voice conversion (VC) trigger that alters timbre and associates it with a target label. Cai et al.\cite{cai2022pbsm} proposed using pitch and timbre as joint triggers for speech backdoor attacks.

\begin{figure}[t]
\centering
\includegraphics[scale=0.38]{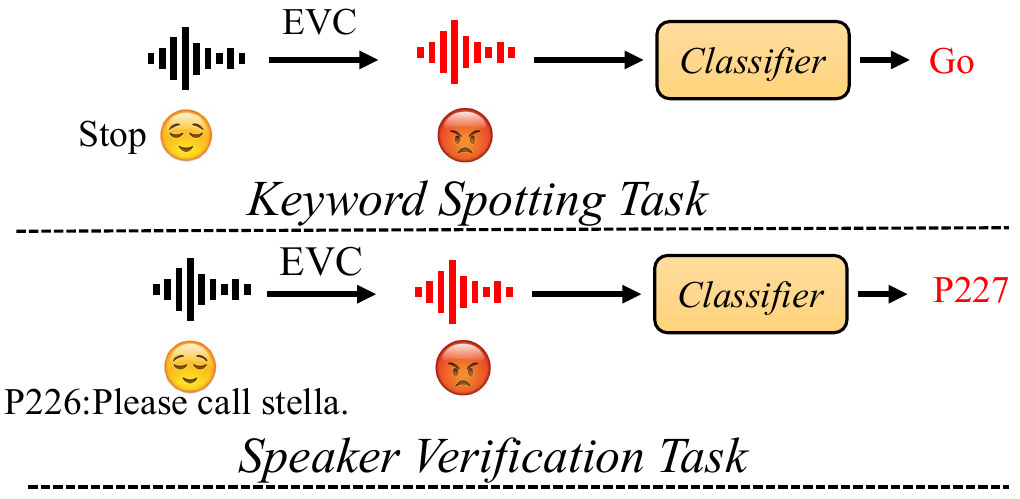}
\caption{The illustration of emotion backdoor attacks. The trigger utilizes the emotional conversion model to generate poisoned samples and the victim model predicts wrongly in two tasks.
}
\label{figure1}
\end{figure}

Based on this, we propose that emotion, a sophisticated composite component of speech formed by rhythm, prosody, and intonation, can also serve as the attack object for speech backdoor attacks. We propose a simple but effective speech backdoor trigger, an emotional voice conversion(EVC) model. The model converts the emotion of speech while preserving other speech components unchanged. We conducted speech backdoor attack experiments on two speech classification tasks, the keyword spotting(KWS) and the speaker verification system(SVs). The victim models were trained on poisoned samples and benign samples, noting that the two sample groups belong to different emotional domains. The results demonstrated that our method shows excellent attack effectiveness and stealthiness on both tasks. 

The main contribution of this work is three-fold:
\begin{itemize}
    \item We reveal that models are suffering from new security risks that come from speech components changing towards the speech utterances.

    \item We proposed the EVC trigger attack paradigm against speech classification models. The attack pipeline is called EmoAttack.

    \item We compared the effectiveness of attacks targeting different emotions. The more significant the emotional difference before and after the trigger, the more influential the attack.
    
\end{itemize}

\section{Background}

\subsection{Backdoor Attacks in Speech Classification}
Considering the characteristics of speech, speech backdoor attacks can be classified into (1) Methods based on the addition of extra noisy speech and perturbation on signals(\textit{\textbf{Noise trigger or Perturbation trigger}})\cite{koffas2023going,koffas2022can,zhai2021backdoor,shi2022audio,liu2022backdoor,liu2022opportunistic,xin2022natural,luo2022practical}. (2) Methods based on the modification of speech components/elements(\textit{\textbf{Element trigger}})\cite{ye2023fake,cai2022vsvc,cai2022pbsm,cai2023towards}. Koffas et.al \cite{koffas2023going} proposed a series of perturbation operations(\textit{e.g.}, pitch shift, reverberation and chorus) to perform digital music effects as a pertubation trigger. The noise trigger also includes the low-volume one-hot-spectrum\cite{zhai2021backdoor} and ultrasonic sounds\cite{koffas2022can}. On the other hand, Ye et.al\cite{ye2023fake,cai2022vsvc} proposed VSVC to treat the timbre as speech backdoor attack trigger. Cai et.al\cite{cai2023towards} also demonstrated that the pitch and timbre triggers could be combined as element triggers for multi-target attacks, which gained excellent attack effectiveness on speech classification models.

\subsection{Emotional Voice Conversion}
Emotional voice conversion (EVC)\cite{zhou2022emotional}  alters the emotional state of speech while retaining its linguistic content and speaker identity, often using acoustic features like mel spectrograms. The EVC model can be expressed as:
\begin{equation}
    x' = EVC(x,e_{t})
\end{equation}
where the $x$ is the speech of source emotion, $x'$ is the converted speech, and the $e_{t}$ is the target emotion category or speech.

\subsection{Speech Classification Tasks.} 

Most speech classification tasks are based on DNN models. The speech classification models include KWS models\cite{simonyan2014very,qin2017dual,gazneli2022end} and SV models\cite{wan2018generalized,desplanques2020ecapa}. The KWS model outputs the speech commands, and the SV models output the speaker embeddings and speaker identification. Both can be trained on signal spectrograms, such as the mel spectrograms and short-time Fourier transform(STFT) spectrograms. The speech classification models can be applied in various fields, such as human-computer interaction, intelligent in-car systems, voice wake-up, and smart robots.

\section{Methodology}

\subsection{Threat Model}
This paper concentrates on poisoning-based backdoor attacks. And there are some basic principles in this scenario. The adversaries can only modify the open-access training dataset to a poisoned dataset. The victim models will be trained on the poisoned dataset, and the user will deploy the models to the working environment. Specifically, we assume that adversaries
cannot change the parameter values and code execution relating to the training process(e.g., loss function, learning schedule, or the resulting model). As shown in Fig. \ref{figure1}, when the attacker alters the emotion of the speech, the KWS model changes its output from 'stop' to 'go,' while the SVs model, initially predicting the speaker as 'p226,' changes to 'p227' after the backdoor was activated.

\subsection{Adversary’s Goals} The attacker's goals are stealthiness, effectiveness, and robustness. Stealthiness means backdoor attacks must evade human and machine detection, with poisoned utterances resembling normal utterances. Effectiveness demands high attack success with minimal poisoning in tests. However, high success rates often require many poisoned samples, reducing stealthiness. Robustness ensures attacks resist simple detection and remain effective against adaptive defences in real-world conditions.

\begin{figure*}[t]
\centering
\includegraphics[scale=0.5]{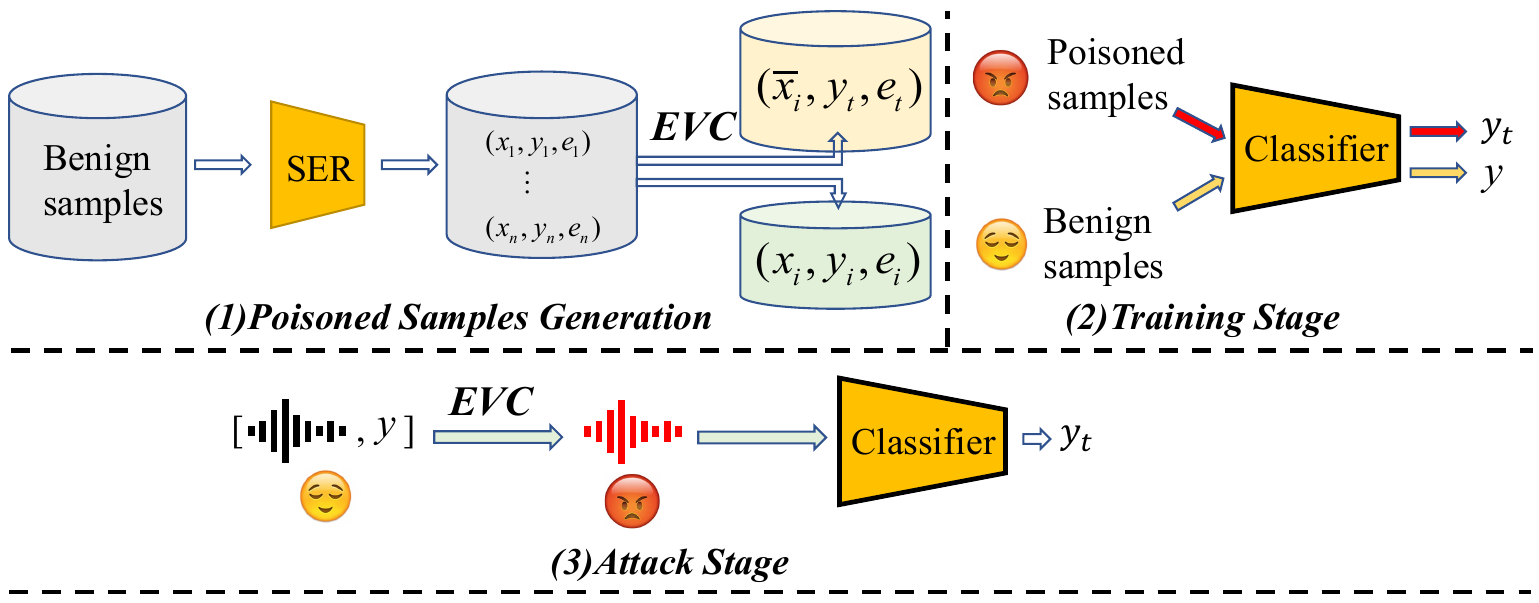}

\caption{The proposed EmoAttack framework.}
\label{figure2}
\end{figure*}


\subsection{Proposed EmoAttack Pipeline}
In this paper, we proposed using emotional voice conversion technologies to product poisoning samples, and the method is called EmoAttack. The attack pipeline includes \textit{(1) Poisoned Samples Generation}, \textit{(2) Training Stage}, and \textit{(3) Attack Stage} as shown in Figure \ref{figure2}. The involved emotions are $\{neutral$, $angry$, $sad$, $surprise$, $happy\}$. We use the Speech Emotion Recognition(SER) model\cite{osman2024ser} to judge the source utterance emotions.

\noindent\textbf{Poisoned Samples Generation.} As shown in Figure \ref{figure2}, the EmoAttack aims to associate the emotional categories of utterances with the target label. It's noted that most samples are classified as neutral emotions by a SER model. We denote a speech sample $x$ with emotion $e$ and label $y$ as $(x,y,e)$. The $N$ samples in the clean dataset $D=\{ (x_{i},y_{i},e_{i}) \}$ are firstly divided into 
two parts: clean training set $D_{t} $, and clean test set $D_{c}$, noting that $D=D_{t}+D_{c}$ and $e_{i}$ denoted the emotion of $i$-th sample. After this, we randomly separated a subset from $D_{t}$ that owned the mostly same emotion, denoted as $D_{tq}$. Based on the top amount of emotion category, we applied the EVC trigger to each sample in $D_{tq}$. We modified their labels to the target label $y_{t}$, resulting in the set $D_{tp} = \{ \overline{x_{i}} = EVC(x_{i},e_{t}), y_{t})\}$, where the $e_{t}$ denoted another target emotion category.



\noindent\textbf{Training Stage.} In the training stage, the backdoor dataset $D_{b}$ is constructed as $D_{b} = (D_{t} - D_{tq} ) + D_{tp}$. The test accuracy would be calculated on the $D_{c}$ set. The classification models were trained on dataset $D_{b}$ and cross-entropy loss.

\noindent\textbf{Attack Stage.} In the attack stage, The attacker can apply the EVC trigger to convert emotions of utterances and activate the backdoor in the victim classifier.

\section{Experiments and Results}

\subsection{Experimental Setting}
\noindent\textbf{Dataset and Models.} We evaluate EmoAttack on the KWS task and SVs task. For the KWS task, we used the Google Speech Commands v2 dataset\cite{warden2018speech}, the victim models are ResNet18\cite{he2016deep}, Attention-LSTM\cite{qin2017dual}, KWS-VIT\cite{berg2021keyword}, EAT-S\cite{gazneli2022end}.  For the SVs task, we used VoxCeleb1\cite{nagrani2017voxceleb} and  TIMIT\cite{garofolo1993darpa}, the victim model are ECAPA-TDNN\cite{desplanques2020ecapa} and SincNet\cite{ravanelli2018speaker}. We partitioned the dataset into training and test sets, the ratio of the training set to the test set is 95 to 5.

\noindent\textbf{Baseline and Trigger Setup.}  We compare EmoAttack with the latest speech backdoor attacks. They are listed as follows. (1) backdoor attack with pixel pattern(BadNets) \cite{gu2019badnets}, (2) position-independent noisy clip backdoor attack(PIBA) \cite{shi2022audio}, (3) dual adaptive backdoor attack(DABA) \cite{liu2022opportunistic}, (4) ultrasonic voice as trigger(Ultrasonic) \cite{koffas2022can}, (5) pitch boosting and sound masking(PBSM) \cite{cai2022pbsm}, and (6) voiceprint selection and voice conversion(VSVC) \cite{cai2022vsvc}. For the EVC trigger, we chose StarGAN-EVC\cite{he2021improved}, which is trained on five emotion domains and can convert the emotion of the input spectrogram to target emotion. We trained the EVC model on the ESD dataset\cite{zhou2022emotional}.

\noindent\textbf{Backdoor Training Setup.} For the KWS task, all victim models were trained by the following set. The batch size is 64, the training epoch is 60, the optimizer is Adam with a learning rate of 1e-4, and all the utterances are segmented or padded into 1 s.  For the SVs task, the models were trained by the following set. The batch size is 128, the training epoch is 100, the optimizer is Adam with a learning rate decreasing from 5e-4 to 1e-4, and all the utterances are segmented or padded into 3 s duration. 

\noindent\textbf{Evaluation Metrics.}
The metrics include attack metrics and trigger metrics. \textit{(1) Attack metrics.} We employ three metrics: Attack Success Rate (ASR), Accuracy Variance (AV), and Poisoned number(PN) to gauge the effectiveness of the backdoor attack. ASR is the attacking behavior on the test dataset. AV refers to the model’s prediction accuracy variance for training before and after the backdoor attacks. Compared with the same datasets, PN directly shows the costs of different triggers for backdoor embedding. \textit{(2) Trigger metrics.} Trigger metrics can prove the trigger's effectiveness.  We use the Mean Opinion Score (MOS) to evaluate speech quality. Furthermore, we selected a SER model to judge the EmoAttack trigger by SER Accuracy, which compares the predicted labels and true labels of attacked emotionally converted samples.

\subsection{Main Results}

\noindent\textbf{Attack Results.} Typically, in backdoor attacks, the ASR gradually approaches $100\%$ only as the number of poisoned samples continuously increases (meaning the poisoning rate keeps rising). Considering ASR and PN results, Table \ref{table:bk_result} shows the experimental configuration parameters for achieving a $99\%$ ASR to facilitate the comparison of method effectiveness. Existing methods and our proposed method both achieve close to $100\%$ ASR under different numbers of poisoned samples. Our proposed method outperforms existing methods in terms of high attack effectiveness due to the lower PN(no more than 100 poisoned samples). It indicates that EmoAttack can successfully implant a backdoor in a speech classification model at a lower cost. From the AV results, we found that our method did not lead to more than $1\%$ AV before and after backdoor attacks, which indicates excellent effectiveness and stealthiness.


\begin{table*}[t]
\centering
\caption{The AV (\%), ASR (\%), and PN of baselines and EmoAttack on two tasks.}
\label{Table2}
\begin{tabular}{lcccccc}
\hline
\multicolumn{1}{l}{}  & \multicolumn{4}{|c}{KWS Task} & \multicolumn{2}{|c}{SVs Task} \\
\hline
\multicolumn{1}{l|}{}        & \multicolumn{1}{c}{ResNet18}   & \multicolumn{1}{c}{Attention-LSTM} & \multicolumn{1}{c}{KWS-VIT} & \multicolumn{1}{c|}{EAT-S}  & \multicolumn{1}{c}{ECAPA-TDNN}   & \multicolumn{1}{c}{SincNet}\\
\hline

\multicolumn{1}{c|}{BadNets} & 0.45/99.97/550                 & 0.32/99.98/550                     & 0.75/99.98/600              & 0.80/99.96/550      & \multicolumn{1}{|c}{0.66/99.85/350}                & 0.70/99.80/400          \\

\multicolumn{1}{c|}{Ultrasonic}  & 2.67/97.82/350 & 2.92/97.68/400 & 3.01/96.92/400     &  2.82/97.25/400    & \multicolumn{1}{|c}{2.05/96.75/400}                 & 2.67/95.12/450  \\

\multicolumn{1}{c|}{PIBA}    & 2.68/94.21/300                & 2.92/93.58/350                     & 3.15/94.62/350             & 3.61/93.59/350           & \multicolumn{1}{|c}{4.16/92.15/300}               & 3.95/93.01/350                     \\
\multicolumn{1}{c|}{DABA}   & 3.65/93.25/450 & 4.21/92.52/400 & 3.91/92.55/450 & 4.55/93.45/450             & \multicolumn{1}{|c}{3.98/94.05/350}                & 4.65/92.81/400               \\

\multicolumn{1}{c|}{PBSM}    & 0.58/99.98/300                 & 0.54/99.88/300                     & 0.72/99.94/350              & 0.66/99.87/350              & \multicolumn{1}{|c}{0.72/99.88/250}                 & 0.64/99.92/300             \\

\multicolumn{1}{c|}{VSVC}                         & 0.51/99.98/250                 & 0.50/99.78/250                     & 0.78/99.92/300             & 0.56/99.93/300             & \multicolumn{1}{|c}{0.72/99.91/250}            & 0.75/99.93/300          \\
\multicolumn{1}{c|}{\textbf{Ours}}    & \textbf{0.55/99.99/50}                  & \textbf{0.62/99.96/50}                      & \textbf{0.72/99.95/60}               & \textbf{0.66/99.97/50}           & \multicolumn{1}{|c}{\textbf{0.75/99.92/60}}                  & \textbf{0.64/99.95/60}        \\
\hline

\end{tabular}
\label{table:bk_result}
\end{table*}

\noindent\textbf{Trigger Evaluation.} Trigger evaluation includes MOS and SER Accuracy, which evaluate whether the poisoned speech samples maintain normal quality. Average MOS is subjective, and SER Accuracy is the objective evaluation by DNN. In the subjective experiment, 10 individuals were invited to participate in an auditory assessment. Each person randomly listened to 30 poisoned samples and the corresponding clean speech samples. They were asked to judge whether the two sentences expressed the same content and whether they sounded normal and gave scores of 0-5. In objective evaluation, we used a SER model to calculate the SER Accuracy on the poisoning test dataset. Specifically, SER Accuracy includes Micro-F1 and Macro-F1 scores. The final results of the evaluation are shown in Table \ref{table:MOS}.

\begin{table}[t]\footnotesize
\centering
\tabcolsep=0.15cm
\caption{The Average MOS and SER Accuracy}
\label{Table3}
\begin{minipage}{\columnwidth}
\begin{tabular}{ccclll}
\hline
\multicolumn{5}{l}{Subjective evaluation: Average MOS}                                                                       &                          \\ \hline
\multicolumn{1}{c|}{\multirow{2}{*}{Average MOS}} & Clean    & BadNets  & \multicolumn{1}{c}{PBSM} & \multicolumn{1}{c}{VSVC} & \multicolumn{1}{c}{Ours} \\
\cline{2-6}
\multicolumn{1}{c|}{}                             & 4.12     & 3.67     & \multicolumn{1}{c}{3.72} & \multicolumn{1}{c}{3.94} & \multicolumn{1}{c}{3.98} \\ \hline
\multicolumn{5}{l}{Objective evaluation: SER Accuracy}                                                                       &                          \\ \hline
\multicolumn{1}{c|}{}                             & Micro-F1 & Macro-F1 &                          &                          &                          \\ \hline
\multicolumn{1}{c|}{Other-to-Neutral}             & 0.7498   & 0.7467   &                          &                          &                          \\
\multicolumn{1}{c|}{Neutral-to-Other}             & 0.7354   & 0.7315  &                          &                          &                          \\ \hline
\end{tabular}
\label{table:MOS}
 \end{minipage}
\end{table}

\begin{figure}[t]
\centering
\includegraphics[scale=0.3]{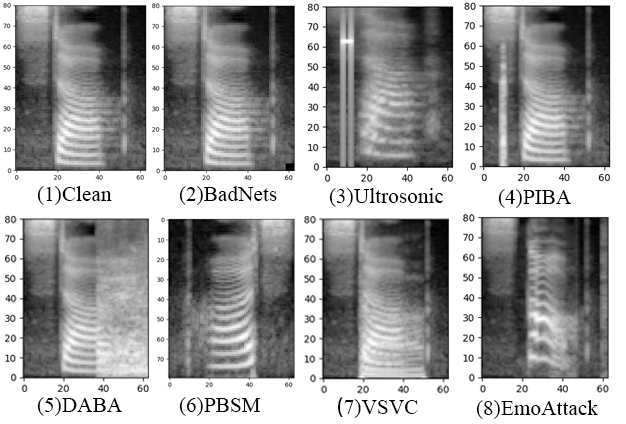}

\caption{Speech backdoor trigger comparison.}
\label{figure3}
\end{figure}

The experimental results in Table \ref{table:MOS} show that our method and VSVC almost do not damage the quality of the speech, so the MOS value is close to the MOS value of ground truth speech. However, methods BadNets and PBSM made detrimental modifications to the spectrogram and fundamental frequency of the speech, resulting in a deterioration in speech quality. Thus, the MOS value is lower than the MOS value of ground truth speech. In the objective experiment, in the clean test dataset, utterances of neutral emotion were converted to ones of other emotion(mostly angry or happy), and utterances of non-neutral emotion were converted to ones of neutral emotion. The F1 values show that the performance of the EVC trigger aligns with the expected effects of the pre-trained model for emotion speech recognition. We also show different speech triggers in Figure \ref{figure3}. The EmoAttack trigger's spectrogram can keep lossless.

\subsection{Ablation Study}
\begin{figure}[ht]
\centering
\includegraphics[height=0.30\linewidth,width=\linewidth]{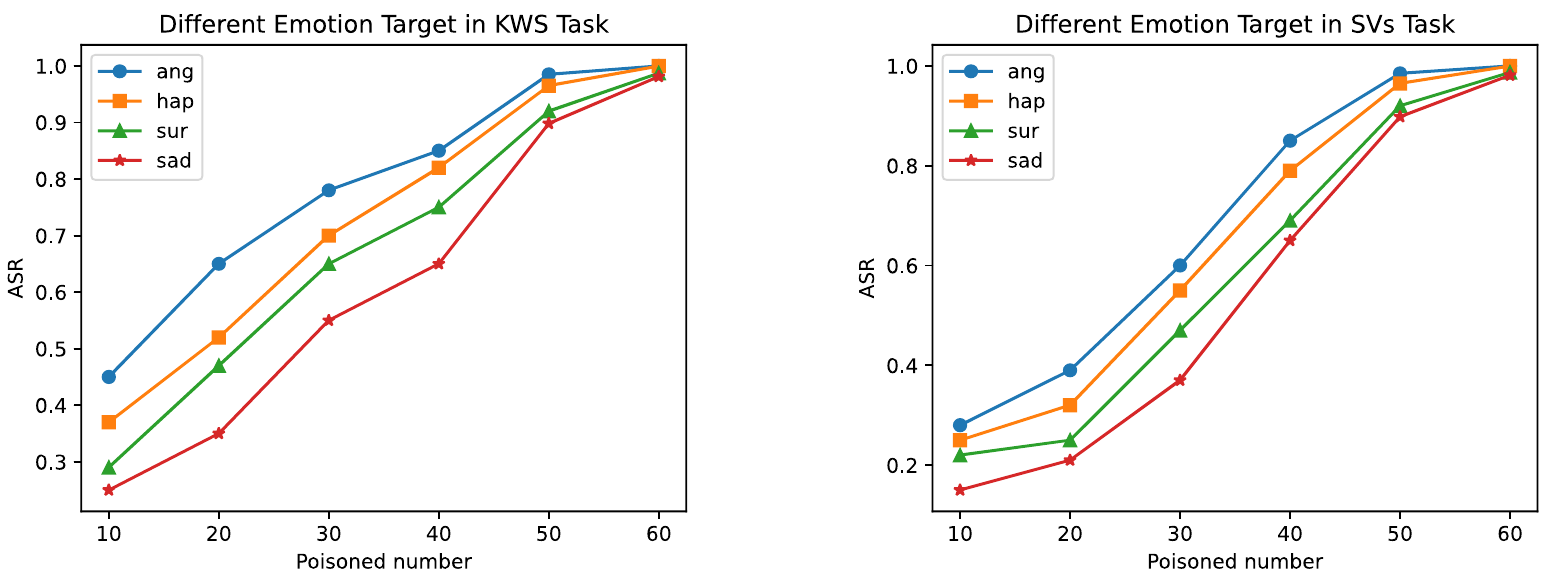}
\caption{ASR with different emotion target configurations.
}
\label{figure4}
\end{figure}

\noindent\textbf{Attack with Different Emotion Target.} Most of the utterances in the dataset are judged as neutral speech. Thus, we chose one of the $\{Angry$, $Happy$, $Sad$, $Surprise\}$ as the target emotion and connected it to a specific target speaker label. We found that intense emotions such as $\{Angry$, $Happy\}$ can achieve the highest ASR most quickly, while the poisoned number gradually reached 60 in two tasks, as shown in Figure \ref{figure4}. In other words, the classification models are more sensitive to these emotions.

\noindent\textbf{Detection Categories by SER Model.} When selecting the samples to be attacked, we used the SER model to determine the maximum voting category of the clean dataset. We selected the samples to be attacked from the data in that category. We found that not doing so may result in unstable or decreased ASR. Therefore, the better attack strategy is to attack the neutral speech.

\section{Conclusion}
This paper analyzes the differences between backdoor attacks in the domains of images and speech. We proposed EmoAttack, a backdoor attack method based on emotional voice conversion. This method preserves speech's linguistic content and timbre characteristics while modifying a higher-level attribute of speech: emotion. After EmoAttack training, the emotional utterances can lead the victim model to produce wrong predictions. We conducted backdoor attack experiments on two speech classification tasks. The experimental results demonstrate the excellent attack effectiveness of the EmoAttack. Additionally, we verified that different emotions as target labels result in varying efficiency of the trigger. The intense emotions gain better results. The proposed method aims to provide insights into backdoor attacks in the speech domain.

\bibliographystyle{IEEEtran}
\bibliography{template}

\end{document}